# How and When the Cassie-Baxter Droplet Starts to Slide on the Textured Surfaces

*Donggyu Kim[1,2], Seunghwa Ryu[1,\*]*

[1]Department of Mechanical Engineering, Korea Advanced Institute of Science and Technology (KAIST), Daejeon 34141, Republic of Korea

**Present Address**

[2]Nature-Inspired Mechanical System Team, Manufacturing Systems Research Division, Korea Institute of Machinery & Materials (KIMM), Daejeon 34103, Republic of Korea

Corresponding author. Fax: +82-42-350-3059

E-mail address: ryush@kaist.ac.kr

**Abstract**

The Cassie-Baxter state droplet has many local energy minima on the textured surface, while the amount of the energy barrier between them can be affected by the gravity. When the droplet cannot find any local energy minimum point on the surface, the droplet starts to slide. Based on the Laplace pressure equation, the shape of a two-dimensional Cassie-Baxter droplet on a textured surface is predicted. Then the stability of the droplet is examined by considering the interference between the liquid and the surface microstructure as well as analyzing the free energy change upon the de-pinning. Afterward, the theoretical analysis is validated against the line-tension based front tracking method simulation (LTM), that seamlessly captures the attachment and detachment between the liquid and the substrate. We answer to the open debates on the sliding research field: (i) Whether the sliding initiates with the front end slip or the rear end slip, and (ii) whether the advancing and receding contact angles measured on the horizontal surface are comparable with the front and rear contact angle of the droplet at the onset of sliding. Additionally, a new droplet translation mechanism promoted by cycle of condensation and evaporation is suggested.

# 1. Introduction

When a droplet is deposited on the lotus leaf, a representative hydrophobic textured surface, it easily slides down from the surface because the air pockets are formed between the droplet and the micro-nano structures on the leaf (i.e. the droplet forms the Cassie-Baxter (CB) wetting state) and thus the effective surface tension between liquid and the textured surface becomes lower than that of the flat surface[1]. Since the sliding characteristic of a droplet is closely related to the extraordinary wetting properties such as self-cleaning[2,3] or anti-icing[4], a plethora of research have been performed to tune the sliding characteristic of a surface with microstructures[5,6].

To explain the sliding phenomena, the conventional theory based on the force balance has been frequently adapted[7,8]. When a droplet is deposited on a tilted surface, because the droplet deforms asymmetrically due to the gravity, the contact angle at the front end becomes larger than the contact angle at the rear end, as depicted in **Fig.1a**. Thus, contact angle difference from the two ends induces the effective surface tension force that pushes the droplet upward, while the gravitational force pulls the droplet downward. The droplet can maintain the shape on the surface if these two forces cancel out each other, but it slides down when the force from surface tension cannot sustain the gravitational force. Therefore, the information on the available ranges of the front and rear contact angles that a droplet can possess on the tilted surface is crucial to predict the sliding condition of the droplet in the framework of conventional wetting theory.

However, several aspects of the droplet sliding phenomenon have not yet been fully understood. To the best of our knowledge, a rigorous explanation on the available ranges of the front and the rear contact angles of the droplet on a tilted surface is absent to date. Hence, the sliding condition of the droplet could not be predicted solely with the surface microstructure information. Additional experimental measurements or some assumptions on the available ranges of the front and rear angles have been necessary for the sliding angle prediction[9-16]. Also, the explanation on the sliding initiation mechanism is still in elusive. Despite the development of high resolution microscopes, it is hard to catch the motion of the droplet tips at the onset of sliding[17]. Hence, it is still open to debate whether the front end or the rear end de-pins ahead when a droplet starts to slide, and indeed both cases were observed in the experiments[5, 13, 18],[19]. In addition, the relationship between the two pairs of contact angles measured on tilted and horizontal surfaces is not fully understood yet; the front and rear angles at the onset of sliding on tilted surface ($\theta_f$ and $\theta_r$ in **Fig. 1a**), and the advancing and receding angles measured on horizontal surface ($\theta_{adv}$ and $\theta_{rec}$ in **Fig. 1b**) [8]. The front/rear angles and the advancing/receding angles have been frequently considered to be identical. However, a doubt on whether the angles really has the same value has been raised [10], and the rigorous comparison has not been performed yet.

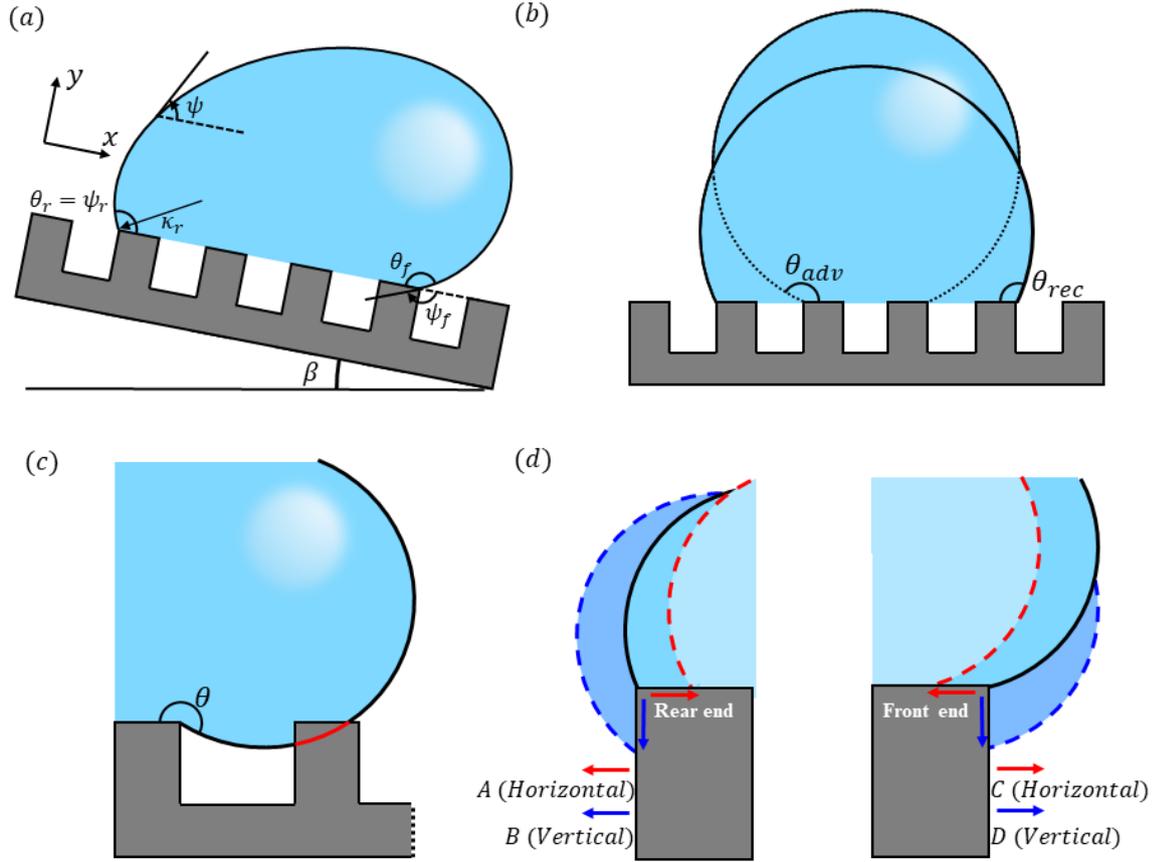

**Figure 1. Sliding of 2D Cassie-Baxter droplet on textured surface**. (a) droplet on tilted surface. On the tilted surface, the droplet has different contact at the front and rear end ($\theta_r$, $\theta_f$). (b) The advancing and receding angle, which are the maximum and the minimum contact angle which the droplet can possess on the horizontal surface. (c, d) The criteria that the droplet must satisfy to maintain the shape. The droplet must not have interference with the surface geometry (c), and both end of the droplet should energetically stable for the perturbation (d).

In this paper, we propose answers to the aforementioned issues on the sliding phenomenon by combining theoretical analysis and the line tension based front tracking modeling (LTM). For the mathematical simplicity, only two-dimensional Cassie-Baxter state on the surface with periodic rectangular surface geometry is considered in this study. First, we present a framework which enables us to obtain the available contact angle ranges at the front and rear end of the droplet on the tilted surface, and predict the sliding angle of the surface solely with the information on the surface microstructure. We first predict the contour of a droplet at each pinning point on the tilted surface according to the numerical calculation based on the Laplace pressure equation[20-22]. Then, the stability of the droplet is determined by examining the geometrical interference between the liquid contour and the textured geometry, as well as the free energy change upon the de-pinning (i.e. slip of the droplet triple junction points) at each end of the droplet (**Fig. 1c, d**) [20, 23, 24]. From the results, we show that the either front or rear tip can move ahead at the onset of sliding, depending on the initial basal area of the droplet. Second, we show that the front and rear angles of the droplet at the onset of sliding are almost identical to the advancing and receding angles of the droplet on horizontal surface for the typical surface microstructures and droplet volume range. Additionally, by analyzing the motion of droplet under condensation and evaporation, we show that the cycle of condensation and evaporation can not only change the shape of the droplet, but also can induce the steady droplet motion. Finally, the theoretical predictions are validated against the line-tension based front-tracking method (LTM), which was recently developed and proposed in our previous study[20], that can seamlessly capture the attachment and detachment between a liquid droplet and textured surface.

## 2. Methods
### 2.1 Stability of the droplet on the tilted surface

It is known that shape of the 2D Cassie-Baxter droplet on the textured surface is decided among the multiple local energy minima points (i.e. pinning points), where the two tips of the droplet is placed at the outer end of the pillars[23]. Therefore, the possible basal length of the droplet ($L$) is not continuous but rather quantized at the pinning points as $L = N(W + G) + W$, and the number of the groove below the droplet $N$ ($N = 0, 1, 2, …$) is used to express the basal length of the droplet, while $W, G$ refer the widths of the step and the groove length of the surface geometry as depicted in the inset of **Fig. 2a**. We first test if the droplet can maintain its shape at the each of the pinning points by examining the interference between the contour of the droplet and the protrusions of the surface. The contour of the droplet is predicted by solving the Laplace pressure equation with Runge-kuta method (See **Methods 2.2**) [20, 22]. For a given droplet volume, it is known that droplet with smaller basal length ($L$) undergoes more severe deformation (i.e. sagging) due to gravity[20]. Therefore, for a given droplet volume and the surface texture, there exists the critical basal length below which interference occurs between the liquid contour and the surface textures. For example, **Fig. 2a** shows the predicted contours of droplets at various pinning points when the droplet with (two-dimensional) volume of $1 \times 10^7 \mu m^2$ is on the tilted textured surface with 25μm of step length W, 100μm of groove length G, 95° of Young's angle $\theta_e$, and 15° of the tilt angle β. When the basal length of a droplet is not large enough, the interference between droplet contour and substrate occurs even with the weaker gravitational effect, (0.08g for N=4, 0.54g for N=14) as depicted with the blue lines, and the interference disappears only if the basal length is increased to the critical value determined by the droplet volume and tilt angle (here, N=21). By employing this method, the minimum value of the basal length of the droplet on the tilted surface can be obtained, in the presence of gravity. We note that such minimum basal length does not exist if the gravitational effect is not taken into account, i.e. droplet sitting on a single pillar is stable[20].

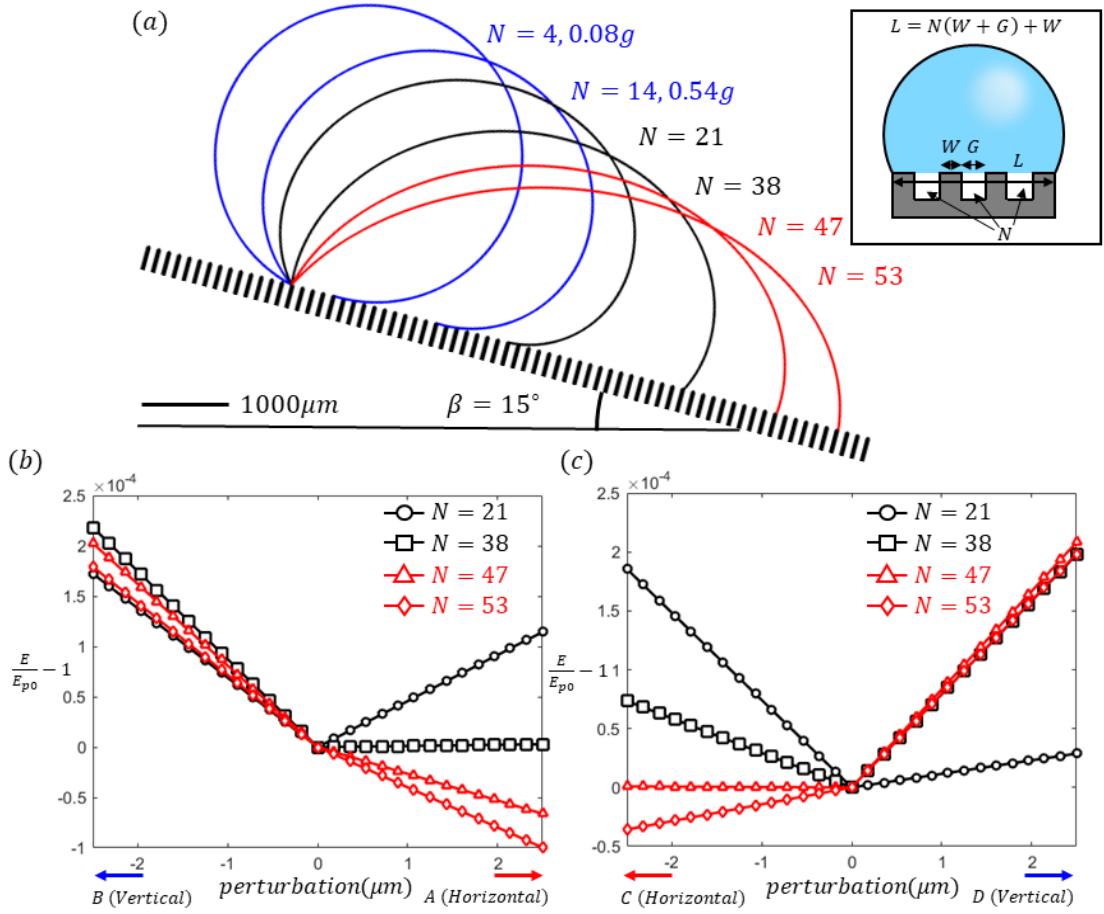

**Figure 2. Examination of the stability of the droplet at each of the pinning points.** (a) The droplet shape at each of the pinning points. The blue lines refer the droplets with geometrical interference with the substrate, the blacks refer the droplets without interference and have stable pinning. The reds refer the droplet without interference but shows unstable pinning. (b) The free energy curve when the perturbation is given at the rear end of the droplet. (c) The free energy curve when the perturbation is given at the front end. The droplet does not have interference and has stable pinning for both the rear and the front end can sustain the shape at the pinning point.

Thereafter, we further examine whether the pinning can be maintained for the droplet which do not have any interference with the protrusions, by analyzing the free energy behavior upon the virtual perturbations on both ends of the droplet in the directions depicted in **Fig. 1d**[20]. The total free energy of the droplet on the tilted surface can be shown as below.

$$E = \gamma_{LV}(A_{LV} - A_{SL}\cos\theta_e) + \rho V g(-x_{cm}\cos\beta + y_{cm}\sin\beta) \quad (1)$$

Where $\gamma_{LV}$ refers the surface tension of the liquid, $A_{LV}, A_{SL}$ refer to the surface area between the liquid and vapor, and the solid and liquid each, $x_{cm}, y_{cm}$ refer to the coordinates of the center of the mass, and $\rho, V, g$ refer the density, volume of the droplet and the gravitational acceleration, respectively. **Figs 2b-c** show how the free energy changes upon the perturbations at the rear and the front ends of the droplets depicted in **Fig 2a**. A (meta-)stable droplet must be at the local energy minimum state with respect to the perturbations for both ends. If one or both ends show the reduction of free energy upon perturbation, the tip of the droplet would slip to lower the free energy upon infinitesimal agitation, and the original state cannot be maintained. When a droplet has appropriate range of basal length ($21 \leq N \leq 38$) as plotted in black in **Fig. 2b** and **2c**, the droplet has its local energy minimum at the end of the protrusions for both front and rear trips, and thus the droplet shape can be maintained. On the other hand, when the basal length is increased to the value corresponds to N >38, the pinning of the rear end becomes unstable, while the front end still shows stable behavior. Therefore, the droplet cannot maintain the shape at the pinning point, and the rear end of the droplet will slip downward to reach the smaller free energy state. When the droplet has longer basal length, as N>47, both the rear and the front end becomes unstable, and the droplet cannot sustain the shape. By conducting the combined geometrical interference and pinning stability analyses, the possible range of the basal length that a stable droplet can possess on the tilted surface is identified ($21 \leq N \leq 38$).

## 2.2. Contour prediction of the droplet on tilted surface

The contour of a CB droplet deformed by gravity at each pinning point on a tilted micro-structured surface is predicted. When the basal length of the droplet is determined, the contour of the droplet can be predicted by solving the Laplace pressure equation. The Laplace pressure equation can be parametrically rewritten as the following set of differential equations[22, 25].

$$\frac{dX}{d\psi} = \frac{cos\psi}{Q}, \frac{dY}{d\psi} = \frac{sin\psi}{Q}, where\ Q = \frac{\rho g}{\gamma_{LV}}(-xsin\beta + ycos\beta) - \kappa_r \quad (2)$$

Here, $X, Y$ refer to the horizontal and vertical coordinates of the droplet contour, respectively, according to the coordinate system visualized in **Fig. 1a**. $\psi$ refers to the angle between the contour and the tilted substrate, $\beta$ refers to the tilt angle, and $\kappa_r$ refers to the local curvature at the rear end of the droplet, as presented in **Fig. 1a**. The differential equations can be solved with the 4$^{th}$ order Runge-Kuta method, when appropriate initial conditions for the contact angles at each of the end $\psi_f$ and $\psi_r$, and the local curvature $\kappa_r$ are given[20]. To find the initial conditions satisfying both given droplet volume and the basal length, the Newton Raphson method is employed with slowly increasing gravitational acceleration from 0 to the real value (9.8 m/s$^2$) (See supporting information).

## 3. Results and discussion

### 3.1 Sliding condition prediction

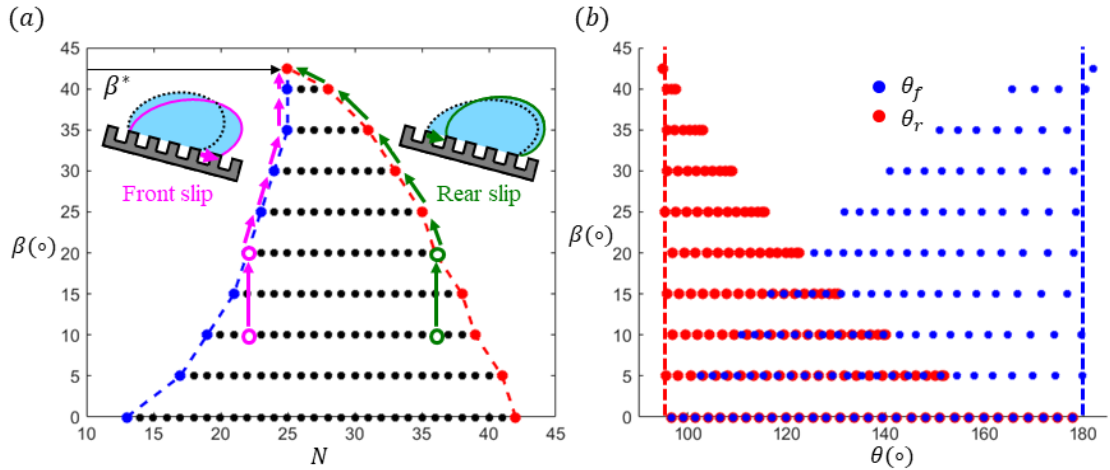

**Figure 3. Stable states of droplet on tilted textured surface**. (a) Number of grooves below the droplet at the stable states about the tilt angle. Each dot refers the stable state of the droplet represented with the groove number N. The number of stable states decreases as the tilt angle $\beta$ increases, and disappears over the $\beta^*$ which is the sliding angle. (b) The front and rear angle of the droplet at each of the stable states. The maximum front angle and the minimum rear angle is maintained at $180°$, and $\theta_e = 95°$, regardless of the tilt angle.

Based on the aforementioned analyses on the possible range of the basal length, we evaluate the possible range of the contact angles at the front and rear ends of the droplet on tilted surface, which also enables us to predict when the droplet starts to slide on the surface. **Figs. 3a-b** show how the available stable states of the droplet represented by the number of grooves below the droplet (**Fig. 3a**) and the front angle $\theta_f$ and rear contact angle $\theta_r$ at each of the stable states (**Fig. 3b**) are changed by the tilt angle of the substrate, $\beta$. When the tilt angle is $0°$, the droplet can maintain the states corresponds to $13 \leq N \leq 42$, in other word, 30 meta-stable states are available. Since the surface is horizontal, the front and rear contact angle has the same value, and the maximum and minimum contact angles becomes the advancing and receding contact angles by its definition. It is noteworthy that the theoretical advancing and receding angles of the Cassie-Baxter droplet on textured surface in the absence of gravity are known to be $180°$ and $\theta_e$ ($95°$ in this example), as revealed elsewhere[20, 23]. If the tilt angle is increased, the number of the available stable states decreases, while the front and rear contact angles deviate from each other, and the possible contact angle ranges for both front and rear end become narrower. Interestingly, although the available number of the stable states changes with the tilt angle, the maximum value of the front contact angle and the minimum value of rear contact angle remain as the advancing and receding contact angles measured on horizontal angle, as presented in blue and red dotted lines. Finally, when the tilt angle reaches to $43°$, only one stable state remains, and the front and rear angles at the state become identical with the advancing and receding contact angles. If the tilt angle is increased beyond the critical value, the droplet cannot find any stable state nor sustain the shape on the surface. Hence, one can conclude that the droplet would slide downward when tilt angle exceeds the critical value, $43°$. With the procedure, we could predict the slide angle $\beta^*$, based on the substrate geometry and Young's angle, without any measurements or hypothesis on the available front and rear angles at the critical state.

**3.2 Tip behavior of the droplet at the onset of sliding**

Based on the framework, we explain the detailed motion of the droplet upon the initiation of the sliding. Despite ever-increasing spatiotemporal resolution of the microscopes, it is still difficult to precisely observe the motion of droplet tips[17], and there have been opposing reports claiming that the sliding initiates from the slip of the front end or the rear end[5, 13, 18, 19]. However, our analysis shows that the sliding can initiate from either the front tip slip or the rear tip slip, which is determined by the initial basal area (which corresponds to the basal length for 2D) of the droplet. For example, if the droplet initially has relatively small basal length (i.e. smaller N) as depicted purple circle in **Fig. 3a**, the droplet can maintain the tip location for the early stage of the tilting. However, when the tilt angle exceeds 20°, the droplet with the initial basal length becomes unstable, and the basal length should increase to settle down on the nearby stable state. Therefore, the front tip slips downward to increase the basal length of the droplet (or the number of grooves below the droplet). Upon further increase of the tilting angle, the basal length extends with the slip of front end until sliding occurs at $\beta = 43°$. On the other hand, for the droplet with initially larger basal length, as depicted in green symbol and arrows, the droplet must reduce the basal length to find the nearby stable state upon tilt angle increase. Thus, the rear end slip occurs ahead of the sliding.

## 3.3 Comparison between advancing and receding angle and the front and rear angle at the onset of sliding

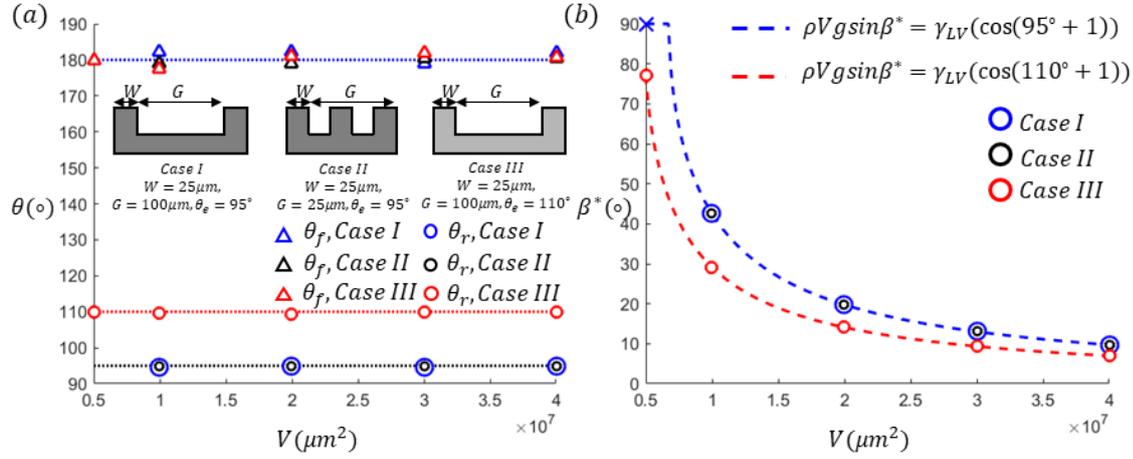

**Figure 4. Sliding condition of droplet on tilted surface.** (a) Front and rear angle of the droplet at critical state. The front angle remains near $180°$, and the rear angle remains near $\theta_e = 95°, 110°$ regardless of substrate geometry or liquid volume. (b) The predicted sliding angle of the droplet at several cases. Since the front and rear angle at the critical state remains at $180°$ and $\theta_e$, the predicted sliding angle matches well with the conventional equation with $\theta_f^* = 180°$ and $\theta_r^* = \theta_e$.

Having analyzed the motion of the droplet onset of the sliding, we turn our attention to another open debate on the equality between the front and rear angles of the droplet at the onset of sliding and the advancing and receding angles measured on the horizontal surface[10]. For the comparison, three different substrate geometry, Case I: $W = 25\mu m, G = 100\mu m, \theta_e = 95°$, Case II: $W = 25\mu m, G = 100\mu m, \theta_e = 110°$, Case III: $W = 25\mu m, G = 25\mu m, \theta_e = 95°$, and a wide range of droplet volume $0.5 \sim 4 \times 10^7 \mu m^2$ are considered. The **Fig. 4a** shows that the front (triangular symbols) and rear angles (circular symbols) of the droplets at the onset of sliding for the three microstructured surfaces are very close to the advancing and receding angles of Cassie-Baxter droplet measured on the same surfaces without tilting, which are known to be $180°$ and Young's angle ($\theta_e$) for typical substrate condition and the wide droplet volume range considered here.

The similarity between the two sets of angles revealed here provides a simpler way to predict the sliding angle of the droplet on the textured surface. The conventional theory[7] predicting the sliding angle based on the force balance uses front and rear contact angle at the onset of sliding ($\theta_f^*$ and $\theta_r^*$ each) as below, where $k$ is the geometric factor which is unity for 2D case[8, 26].

$$\rho V g \sin\beta^* = k\gamma_{LV}(\cos\theta_r^* - \cos\theta_f^*) \qquad (3)$$

Therefore, by plugging $180°$ and $\theta_e$ into the $\theta_f^*$ and $\theta_r^*$, the sliding angle can be simply predicted with the formula, $\rho V g \sin\beta^* = \gamma_{LV}(\cos\theta_e + 1)$. **Fig. 4b** shows the predicted sliding angle of the surface for various droplet volume with the framework proposed in this study (depicted in symbols), and with the equation (3) (depicted in dotted lines). Two results show a great agreement with each other for the various surface microstructures and the droplet volume ranges with varying sliding angles (critical tilt angle) in the range of $10° - 90°$.

We then discuss the multiple factors affecting the sliding phenomenon based on the analyses. First, the effect of surface microstructure can be examined from the comparison between the case I (blue symbols) and case II (black symbols). As presented in **Fig.4**, they show almost same $\theta_f^*$ and $\theta_r^*$, as well as the sliding angle (i.e. critical tile angle) for the entire range of droplet volume. However, it does not imply that the surface texture geometry does not affect the sliding phenomena because the analysis is based on the two-dimensional droplet. For the three-dimensional droplet, the surface roughness perpendicular to the sliding direction reduces the effective surface tension, which can ultimately decrease the effective Young's angle of the surface. Still, one can suspect that the roughness along the sliding direction would have a limited effect. Second, the effect of the surface chemistry can be examined by comparing Cases I and III. For both cases, the front angles at the onset of sliding are maintained near $180°$, while the rear angles are at the Young's angle of each substrate. Therefore, the substrate made of a material possessing higher Young's angle has smaller difference between the front and rear angles at the onset of sliding, which implies easier sliding.

**3.4 Droplet motion promoted by cycle of condensation and evaporation**

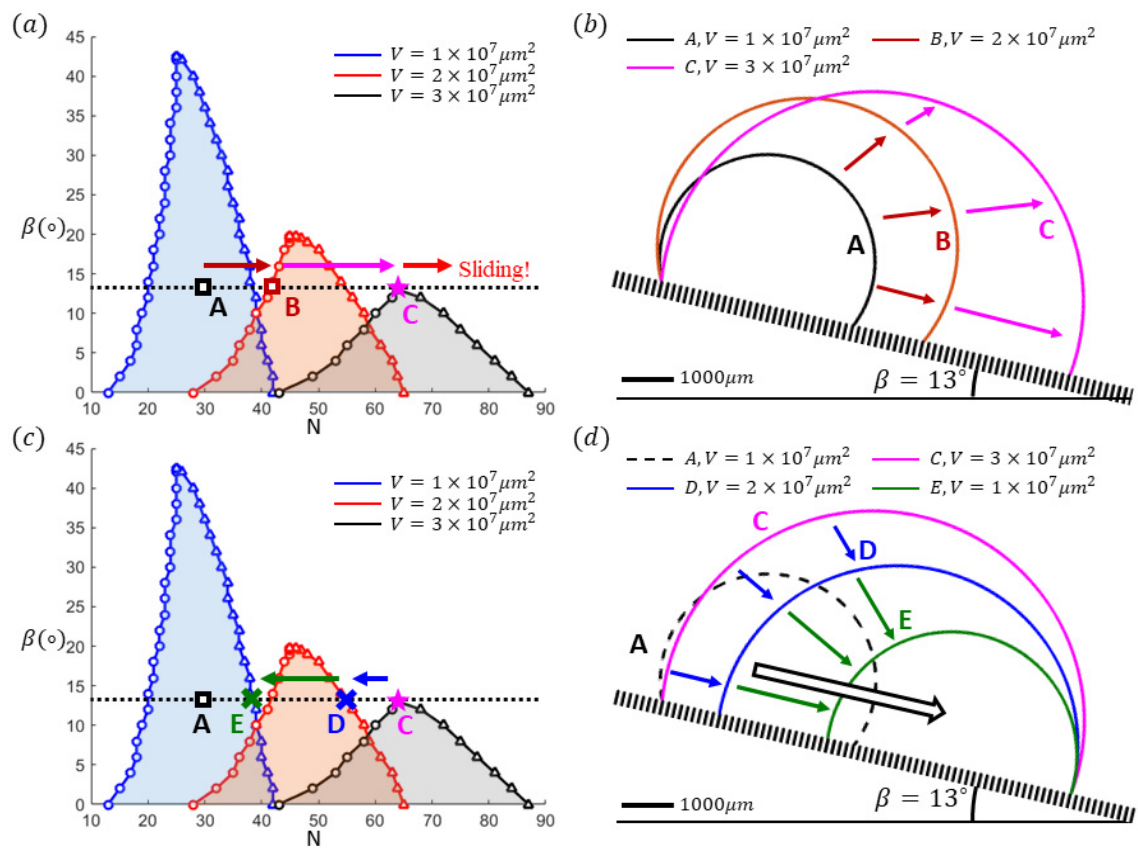

**Figure 5. Tip behavior of the droplet under condensation and evaporation.** (a) Stable states of droplet represented with the groove number N under condensation. (b) Droplet shape change under condensation. The front end slip is occurred before sliding to increase the groove number N. (c) Stable stats of droplet represented with the groove number under evaporation. (d) Droplet shape change under evaporation. The rear end slip is occurred before sliding to decrease N and the droplet can steadily move on the surface by a cycle of condensation and evaporation (**A → E**).

Additionally, we analyze the behavior of a droplet under the cyclic condensation and evaporation, in order to search for the practical applications from the findings, related to water locomotion device, water harvesting, and etc. We find that the cyclic condensation and evaporation is able to not only change the shape of droplet but also induce the steady translational motion of the droplet. For a test case, the textured substrate with $W = 25\mu m, G = 100\mu m$, and $\beta = 13°$ was prepared. **Figs. 5a,c** shows the available stable states of the droplet represented by the groove number N, for three different volumes, $V = 1 \times 10^7 \mu m^2, 2 \times 10^7 \mu m^2$, and $3 \times 10^7 \mu m^2$. Suppose that a droplet with volume of $1 \times 10^7 \mu m^2$ is initially located at the position A. If the volume of the droplet is increased to $2 \times 10^7 \mu m^2$ with condensation, the original state becomes unstable and the basal length of the droplet must be increased to stabilize the droplet. Thus, the front end slip occurs, as shown in **Fig. 5b** (A→B). If the continued condensation is occurred during which the droplet volume reaches to $3 \times 10^7 \mu m^2$ (B→C), only one stable state becomes available at the tilt angle, and the irreversible sliding would be occurred for further condensation. Therefore, in the process of condensation, the sliding is initiated with the front end slip.

On the other hand, **Figure 5c** and **5d** depict the behavior of the droplet during the evaporation process. In the process, the basal length or the number of grooves below the droplet (N) should be decreased for the droplet to find the nearby stable state, thus the rear end slip occurs (C → D → E). Interestingly, a droplet that underwent a full cycle of condensation and evaporation must have the longest basal length among those of possible stable states. Thus, the droplet shape is modified by the condensation and evaporation cycle from the initial shape and detailed process is visualized in the **Fig. 5d** (A → E). Also, as depicted in **Fig 5d**, because the front end slip is occurred for the condensation process and the rear end slip is occurred for the evaporation process, the droplet moves downward on the tilted surface through the condensation and evaporation cycle. For validation, the phenomenon is also modelled with the Line tension based front tracking method (LTM) for smaller scale and larger gravitational effect, ( $W = 25\mu m, G = 100\mu m, \theta_e = 140°, g = \frac{98m}{s^2}, \beta = 10°, V = 3 \times 10^5 \sim 7 \times 10^5 \mu m^2$ ) , and the result is in the **supplementary video**.

The movement of the droplet is distinct from the sliding phenomenon of the droplet because the droplet is in semi-steady state, which refers the droplet is always in the stable state during the movement, while the intermittent sliding happens when the droplet is not capable to find the stable state upon volume change. Therefore, the phenomenon can be further utilized for precise transportation of a droplet.

### 3.5 Comparison with the Line tension based front tracking method (LTM)

Finally, the theoretical analysis is compared with the virtual experiments based on the line tension based front tracking method (LTM) [20] because of the difficulty performing experiments on two dimensional CB droplet. As shown in our previous study, LTM can capture the attachment and detachment between the liquid surface and the textured geometry. LTM discretize the contour with finite number of segments, and consider that the segments near the substrate belong to the liquid-solid interface and the other segments belong to the liquid-gas interface. Utilizing the augmented Lagrange method, the simulation updates the segments along the droplet contour in a way to minimize the total free energy while satisfying constraints on constant droplet volume. The detailed description on the simulation is described in the **supporting information**. For validation, the sliding behavior of a case considered in the prior theoretical analysis (the droplet volume of $1 \times 10^7 \mu m^2$ on the textured surface $W = 25 \mu m, G = 100 \mu m, \theta_e = 95°$) is modelled with LTM. We prepare two droplets with different initial shapes, which the configurations at the advancing and receding states (i.e. the states with the minimum and maximum basal length each among the available stable states) on the horizontal surface, and analyze the shape and tip behavior of the droplet for the gradual increase of the tilt angle.

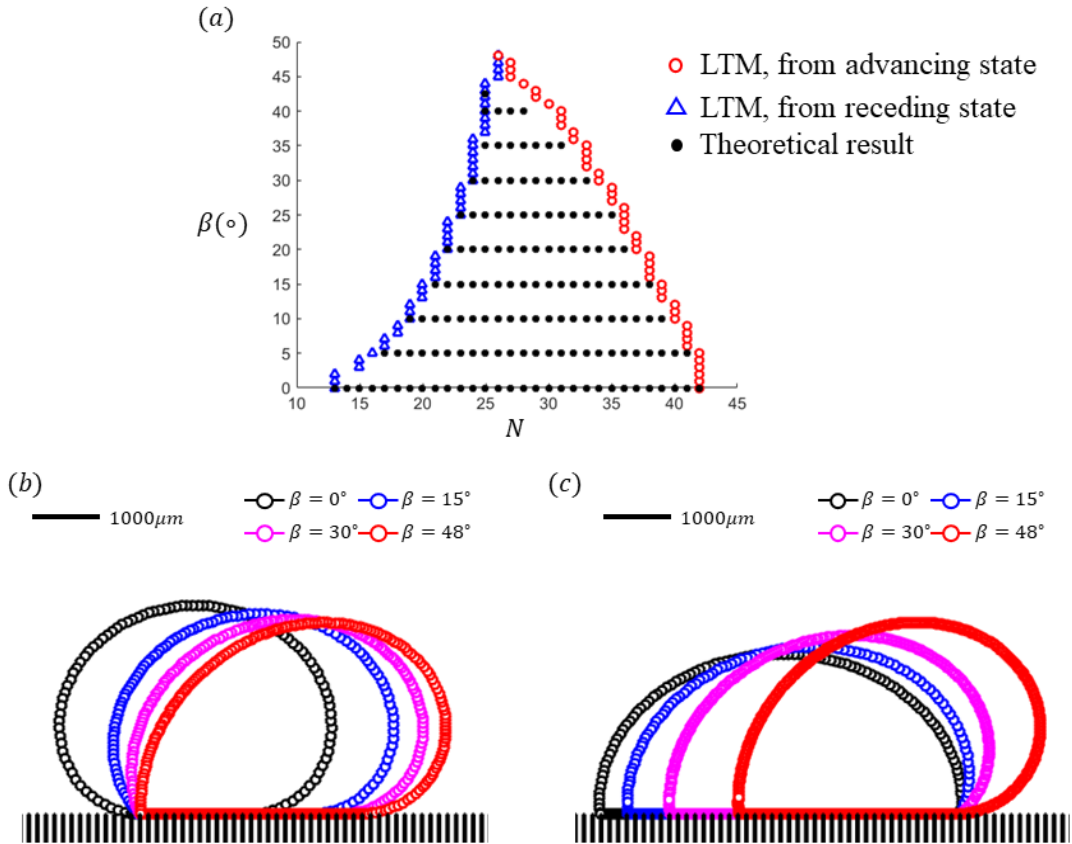

**Figure 6. Line tension based front tracking method results of droplet on tilted textured surface.** (a) Relation between number of grooves below the droplet (N) and the tilt angle ($\beta$) for stable states on the tilted surface. (b) The configuration of droplet from advancing condition, when the droplet volume $1 \times 10^7 \mu m^2$ is on the surface with $W = 25\mu m$, $G = 100\mu m$, $\theta_e = 95°$. As the tilt angle is increased, The front end of the droplet moves while the rear end is pinned. (c) The configuration of droplet from receding condition. The rear end moves as the tilt angle is increased, while the other end is pinned.

**Fig. 6a** shows the comparison of the droplet behaviors before the sliding, predicted from the LTM and the theoretical analysis. The blue triangular and red circular symbols refer to droplet shapes from the LTM, initially positioned at the advancing (i.e. minimum N state) and receding state (i.e. maximum N state) each, while the black dots refer the possible stable states of the droplet from the theoretical analysis. When the tilt angle is increased for the droplet which was initially positioned at the advancing state, the front end slip is occurred before the sliding while the rear end is still pinned as shown in **Fig.6b**, thus the basal length (or the groove number below the droplet) increases, showing good agreement with the trend of the maximum

N states among the possible states predicted from the theoretical results. On the other hand, for the droplet initially at the receding state, the rear end slip is occurred and the basal length is decreased as the tilt angle is increased following the similar path with the maximum N states from the theoretical result. When the tilt angle reaches to 48° (red in **Fig. 6b, c**), the shape of the both droplets initiated from the two different states becomes almost identical, and no more appropriate stable state is found for tilt angle above the critical value. While the exact value of the sliding angle from the LTM is deviated slightly with the difference of ~5° because of the protrusion edge effect of the LTM (refer the details in the Supporting Information), the trends from LTM and the theoretical analysis still shows good agreement.

## 4. Conclusion

In summary, we present the new framework of theoretical model which can explain and predict the sliding phenomenon of the droplet on textured surface. The examination of the stability of the droplets positioned at each of the possible pinning points on the tilted surface is conducted with the geometrical interference test and the free energy analysis on the virtual perturbation on each end of the droplet. As a result, the possible range of front and rear angle of the droplet on the tilted surface is decided and the sliding condition of the surface is predicted without any measurement or a prior assumption on the front and rear angle. Thereafter, the tip movement of the droplet before the sliding is analyzed, and it is shown the sliding can occur from either the front end slip or the rear end slip, and it is decided by the initial basal area of the droplet. Also, the front and rear contact angle at the onset of the sliding is compared with the advancing and receding angle measured on the horizontal surface, and they had very similar value for typical substrate condition and wide range of liquid volume. Finally, the movement of the droplet under the cyclic condensation and evaporation process is analyzed, and it is uncovered that the cycle of condensation and evaporation can steadily move the droplet on the surface, which can be further utilized for precise droplet transportation. The presented work provides deeper physical understanding on the sliding and the tip behavior of the droplet on tilted surface, can be further applied to design surface texture with desired wetting properties

**Acknowledgements**

Supporting Information

# How and When the Cassie-Baxter Droplet Starts to Slide on the Textured Surfaces

**Supplementary Note 1: Prediction of droplet contour on tilted textured surface**

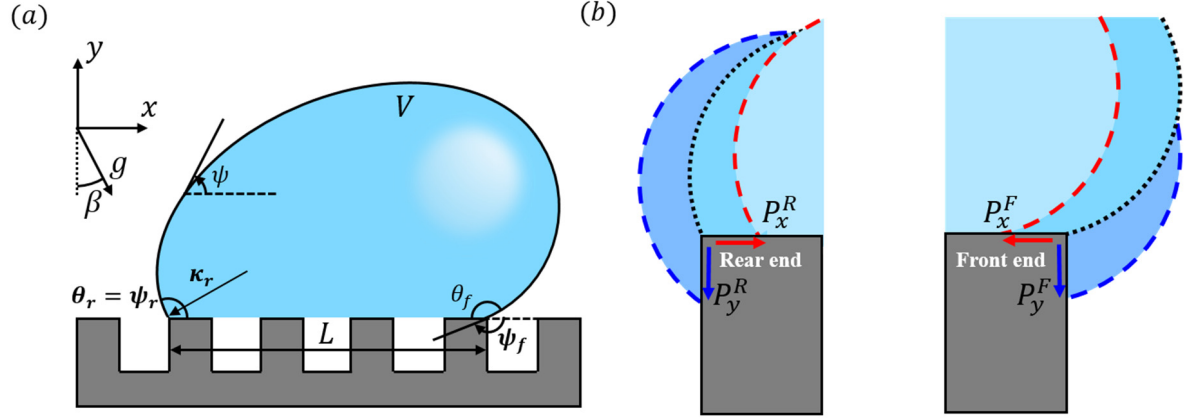

**Figure S1. Modelling of droplet with gravitational effect.** (a) Droplet on tilted textured surface (b) Virtual perturbation on rear end ($P_x^R, P_y^R$) and front end ($P_x^F, P_y^F$)

The contour of the droplet ($X, Y$) on tilted surface follows the Laplace equation, which can be parametrized as follow differential equation[1],

$$\frac{dX}{d\psi} = \frac{\cos\psi}{Q}, \frac{dY}{d\psi} = \frac{\sin\psi}{Q}, \text{where } Q = \frac{\rho g}{\gamma_{LV}}(-x\sin\beta + y\cos\beta) - \kappa_r \quad \text{(S1)}$$

The differential equation can be solved with the 4th order Runge-kuta method using appropriate initial conditions of $\psi_r, \psi_f, \kappa_r$, where $\psi_r$ and $\psi_f$ are the angles between the liquid surface and the surface at of two ends of the droplet, and $\kappa_r$ is the local curvature of the droplet at the rear end as depicted in **Fig.S1**.

Thereafter, the set of initial condition $\mathbf{S} = (\psi_A, \psi_B, \kappa_A)$, which constructs the droplet that satisfies the given liquid volume and the tip location is found with the Newton Raphson method. First, the initial condition satisfying the given droplet condition (liquid volume and the tip location) at zero gravity was calculated from the geometrical relation. Then, the error function $f(\mathbf{S})$, which is the difference between the given droplet condition and the condition

from the constructed droplet with $S$ is calculated for non-zero gravity

$$f(S) = [V(S) - V_0, L(S) - L_0, D_y(S) - D_{y0}]^T \tag{S2}$$

Here, $V_0, L_0, D_{y0}$ each refers to the given volume, basal length, and the height (in y direction) difference between each end of the droplet, and $V(S), L(S), D_y(S)$ each refers to the corresponding values from the droplet constructed with the initial condition $S$.

Once the error function is computed, the Newton Raphson method is employed to find the appropriate initial condition until the error function becomes small enough.

$$S_{i+1} = S_i - J(S_i)^{-1}f(S_i), \quad J(S) = \begin{bmatrix} \frac{\partial f_1(S)}{\partial S_1} & \frac{\partial f_1(S)}{\partial S_2} & \frac{\partial f_1(S)}{\partial S_3} \\ \frac{\partial f_2(S)}{\partial S_1} & \frac{\partial f_2(S)}{\partial S_2} & \frac{\partial f_2(S)}{\partial S_3} \\ \frac{\partial f_3(S)}{\partial S_1} & \frac{\partial f_3(S)}{\partial S_2} & \frac{\partial f_3(S)}{\partial S_3} \end{bmatrix} \tag{S3}$$

Here, $J(S)$ is the Jacobian matrix, and $S_1, S_2, S_3, f_1, f_2, f_3$ each refers to the element of $S$ and $f(S)$ each. The process is repeated for gradually increasing gravity. For the droplet which shows interference between the liquid and the surface protrusion before the gravity reaches to the real value, further calculation was not made and the droplet was assumed to also have interference when the gravity becomes the real value.

When the virtual perturbation is exerted on the tip of the droplet to analyze the pinning stability, the given condition of the droplet is modified with the amount of the perturbation on each of the end.

$$L_P = L_0 - (P_x^R + P_x^F), D_{yp} = D_{y0} - (P_y^F - P_y^R) \tag{S4}$$

Here, $L_P, D_{yP}$ refers to the modified basal length and height difference between the tip of the droplet and $P_x^R, P_y^R, P_x^F, P_y^F$ each refers to the horizontal and vertical perturbations on the front and rear end of the droplet each as **Fig.S1b.** Thereafter, similar procedure is repeated to predict the contour of the droplet.

**Supplementary Note 2: Line tension based front-tracking method (LTM) for 2D CB droplet on titled textured surface**

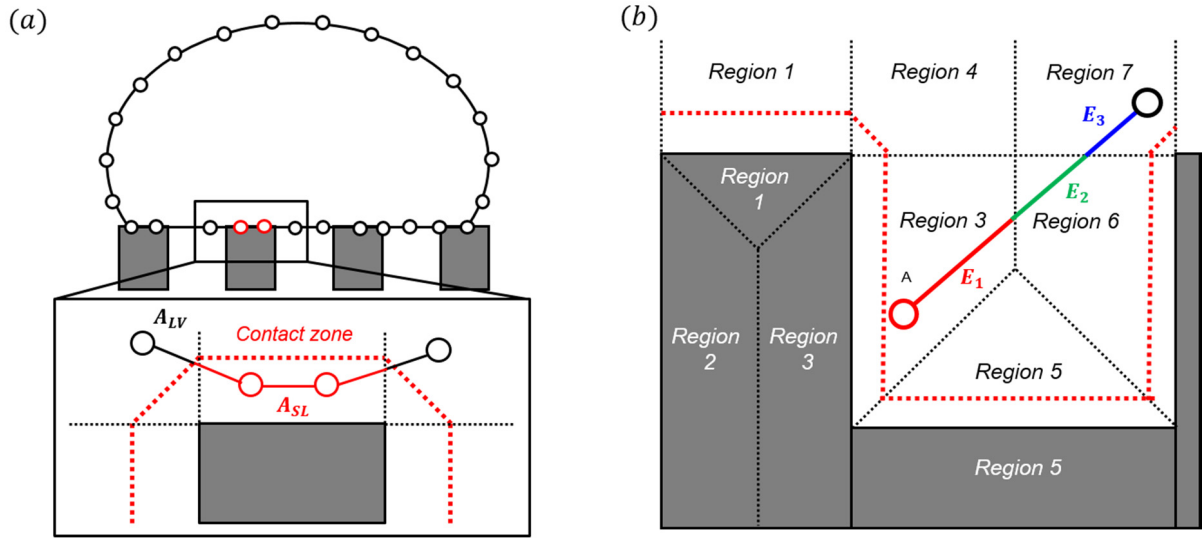

**Figure S2. Line tension based front-tracking method**. (a) Elements constructing the droplet. The elements near the protrusion are considered as the surface between liquid and the surface ($A_{SL}$) and the others are considered as the surface between the liquid and the vapor ($A_{LV}$) (b) Regions of the textured surface. The periodicity of the textured surface is divided into 7 regions, and the surface energy of the element is computed by adding the energies calculated from each regions.

The presented theoretical results are validated after the line tension based front-tracking method (LTM), which recently developed and presented elsewhere [2]. The LTM models the droplet with finite number of elements and relocate the elements in the direction of minimizing the total free energy while preserving the constraints on the droplet, as other simulation[3]. However, the numerical method is capable of modelling the attachment and detachment phenomenon between the liquid and the surface, which is very important in analyzing the wettability of the rough surface. Distinct from the usual simulation which the surface type of the element is fixed during the iteration, the surface type of the elements are continuously changed depending on the position of the elements in LTM simulation. As shown **Fig.S2a,** the LTM assumes element near the protrusions (Contact zone, the section inside the red dotted in **Fig.S2a**) as the surface between liquid and the substrate, and the others as the surface between the liquid and the vapor. And because the boundary of the section is smoothened with the hyper tangent function, the element can freely move across the boundary to minimize the total free energy of the droplet.

Using the changeable surface type of the element, the wettability of the surface with the rectangular protrusion is modelled. The periodicity of the surface with the protrusion is divided into 7 regions, and the surface energy of the element crosses more than one region is calculated by adding the surface energies computed at each of the region as depicted in **Fig.S2b**. Since the LTM assumes finite thickness of the contact zone, the LTM inevitably has fillet at the end of the protrusions as shown in region 4 and 7 of **Fig. S2b**. The edge effect becomes smaller as the thickness of the contact zone becomes thinner, however the computational time also increases with it. In the simulation shown in the manuscript, the $0.2\mu m$ thickness of the contact zone is employed, but the simulation result shows considerable difference in the number of stable states that the droplet can possess, as the tilt angle of the surface approaches to the sliding angle of the surface. The results presents the possibility that the fillets generated during the manufacture of the protrusion may affect the sliding performance of the surface